\documentclass[10pt]{article}
\usepackage{graphicx}
\usepackage{amsmath}
\usepackage{multirow}
\def\Title#1{\begin{center} {\Large #1 } \end{center}}
\def\Author#1{\begin{center}{ \sc #1} \end{center}}
\def\Address#1{\begin{center}{ \it #1} \end{center}}

\newcommand\pubblock{\rightline{\begin{tabular}{l} Proceedings of the Fifth Annual LHCP\\ \pubnumber\\
         \pubdate  \end{tabular}}}

\newenvironment{Abstract}{\begin{quotation} \begin{center} 
             \large ABSTRACT \end{center}\bigskip 
      \begin{center}\begin{large}}{\end{large}\end{center} \end{quotation}}

\newenvironment{Presented}{\begin{quotation} \begin{center} 
             PRESENTED AT\end{center}\bigskip 
      \begin{center}\begin{large}}{\end{large}\end{center} \end{quotation}}

\def\beq{\begin{equation}}
\def\eeq#1{\label{#1}\end{equation}}
\def\eeqn{\end{equation}}

\def\beqa{\begin{eqnarray}}
\def\eeqa#1{\label{#1}\end{eqnarray}}
\def\eeqan{\end{eqnarray}}

\let\bar=\overbar

\def\Dslash{\not{\hbox{\kern-4pt $D$}}}
\def\dslash{\not{\hbox{\kern-2pt $\del$}}}

\def\msb{{\bar{\ssstyle M \kern -1pt S}}}

\usepackage{color}

\newcommand\pubnumber{ ATL-PHYS-PROC-2017-110 }
\newcommand\pubdate{\today}

\def\affiliation{
On behalf of the ATLAS Collaboration, \\
Duke University, USA\\
Tsung-Dao Lee Institute, China\\
Shanghai Jiao Tong University, China}

\begin{document}

% large size for the first page
\large
\begin{titlepage}
\pubblock

%% Change the title, name, abstract
%% Title 
\vfill
\Title{Studies of $Z\gamma$ electroweak production in association with a high-mass di-jet system in $pp$ collisions at $\sqrt{s}$ = 8 TeV with the ATLAS detector}
\vfill

%  if you need to add the support use this, fill the \support definition above. 
%   \Author{ FIRSTNAME LASTNAME \support }
\Author{ Shu Li  }
\Address{\affiliation}
\vfill
\begin{Abstract}

The production of the $Z$ boson and a photon in association with two high-mass di-jets is studied,
using 20.2 fb$^{-1}$ of proton-–proton collision data at a centre-of-mass energy of $\sqrt{s}$ = 8 TeV
recorded with the ATLAS detector in 2012 at the Large Hadron Collider. The measurement focuses on 
fully leptonic decayed $Z$ boson for both charged leptons ($\ell = e/\mu$) and neutrinos.
The measured cross-sections are for both electroweak and total $pp \to Z\gamma j j$ (including QCD induced components)
are extracted in two fiducial regions with different sensitivities to
electroweak-only processes. Quartic couplings of vector bosons are studied
in regions of phase space with an enhanced contribution from pure electroweak
production, sensitive to vector boson scattering processes $VV \to Z\gamma$.
No deviations from  Standard Model predictions are observed and constraints are placed on
anomalous couplings parameterized by higher-dimensional operators using effective field
theory.

\end{Abstract}
\vfill

% DO NOT CHANGE 
\begin{Presented}
The Fifth Annual Conference\\
 on Large Hadron Collider Physics \\
Shanghai Jiao Tong University, Shanghai, China\\ 
May 15-20, 2017
\end{Presented}
\vfill
\end{titlepage}
\def\thefootnote{\fnsymbol{footnote}}
\setcounter{footnote}{0}
%

% normal size for the rest
\normalsize 

%% Your paper should be entered below. 

\section{Introduction and Physics Motivations}

The measurement of the pure electroweak processes in diboson plus two jets final states is an important probe of electroweak symmetry breaking (EWSB),
especially through the vector boson scattering (VBS) topology. Without the presence of the Higgs boson, the longuitudinally polarized VBS processes' amplitudes
increases as a function of the center-of-mass energy and violates unitarity at sufficiently high energies. So the Standard Model (SM) VBS also opens the window to
probe the Higgs mechanism in unitarization of such quartic coupling processes.

This proceeding presents the first results in ATLAS of such pure electroweak processes in $Z\gamma+jj$ final states and probe for the first time the beyond Standard Model (BSM)
anomalous quartic gauge couplings (aQGC) via probing the neutral gauge coupling vertices of $ZZZ\gamma$/$Z\gamma Z\gamma$/$\gamma \gamma Z\gamma$, which are prohibitted in SM at tree-level,
alongside the charged vertex $WWZ\gamma$.

The measurements employs both charged lepton and neutrino final states of $Z$ boson decays.
The fiducial cross-sections are measured for the two channels separately and the upper limits on the anomalous couplings are set with the two channels combined.

\begin{figure}[h!]
\small
  \begin{center}
  \includegraphics[width=0.33\columnwidth]{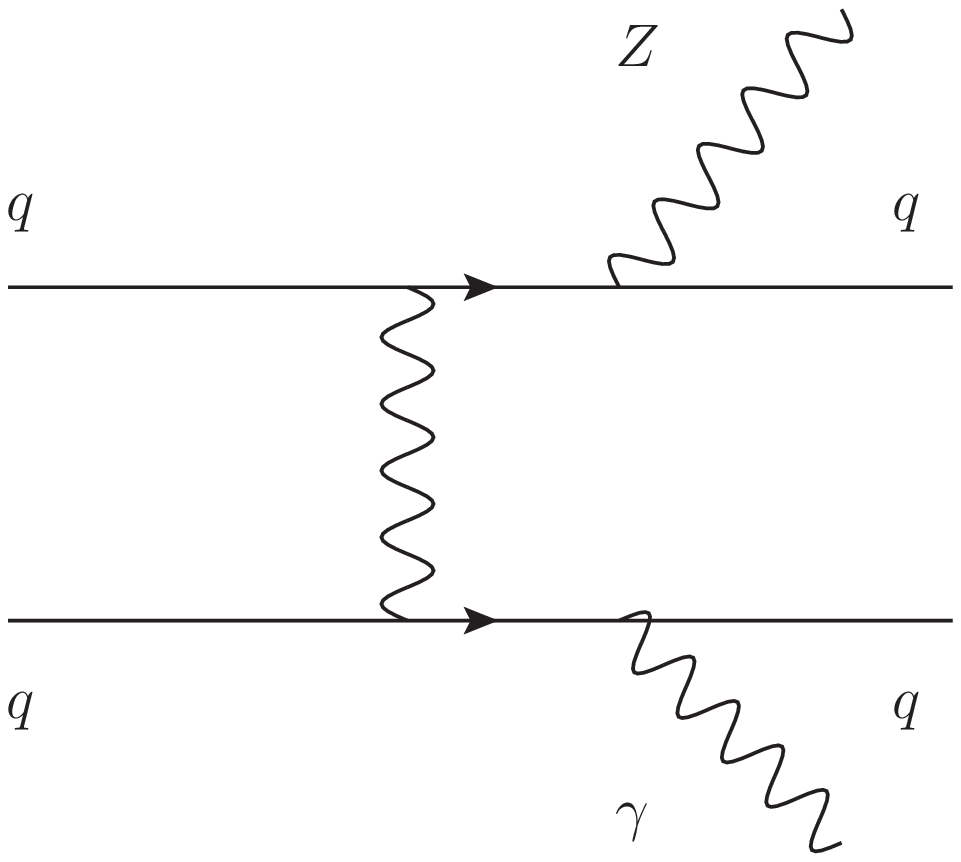}
  \hspace{2cm}
  \includegraphics[width=0.33\columnwidth]{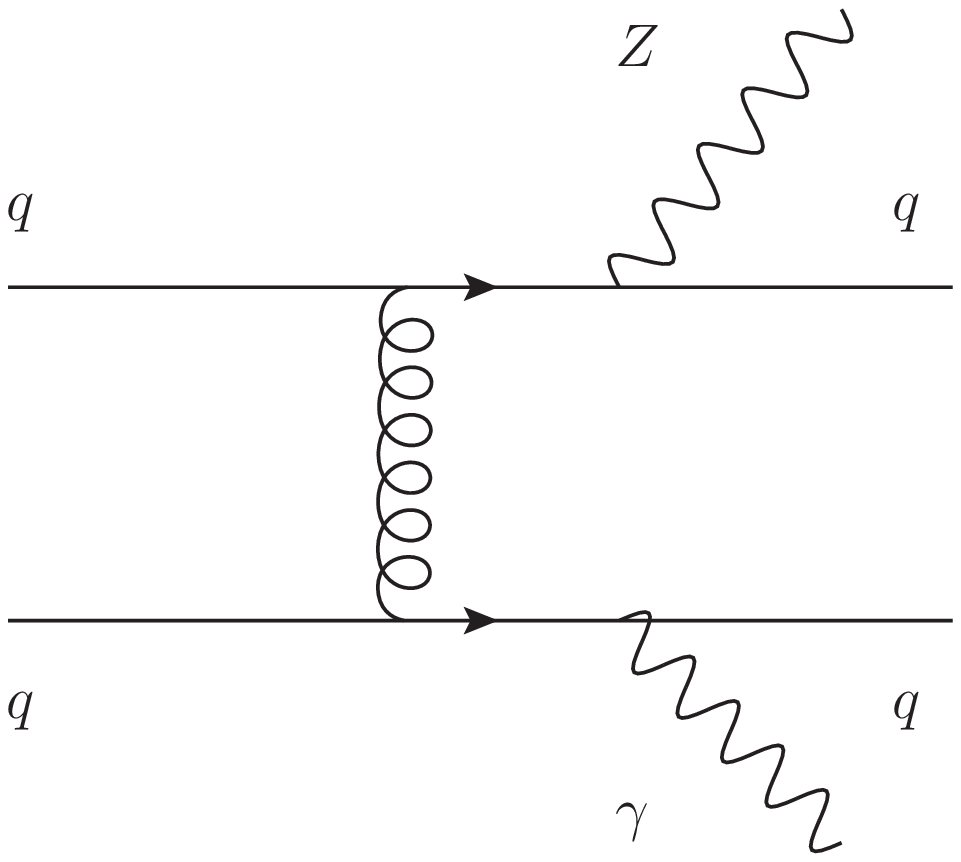} \\
  \includegraphics[width=0.33\columnwidth]{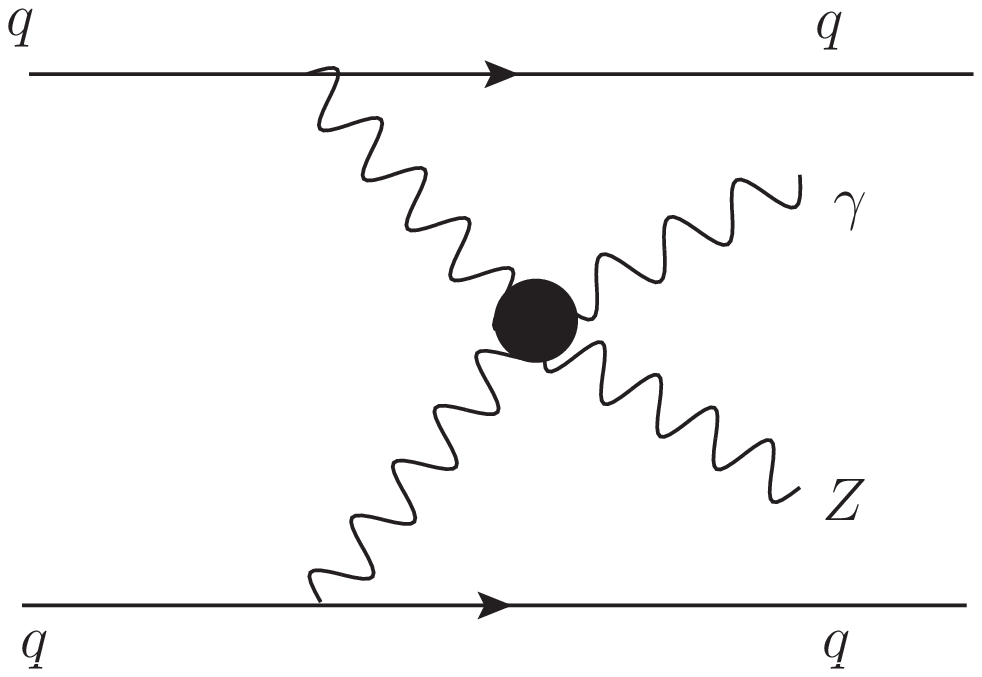}
  \hspace{2cm}
  \includegraphics[width=0.33\columnwidth]{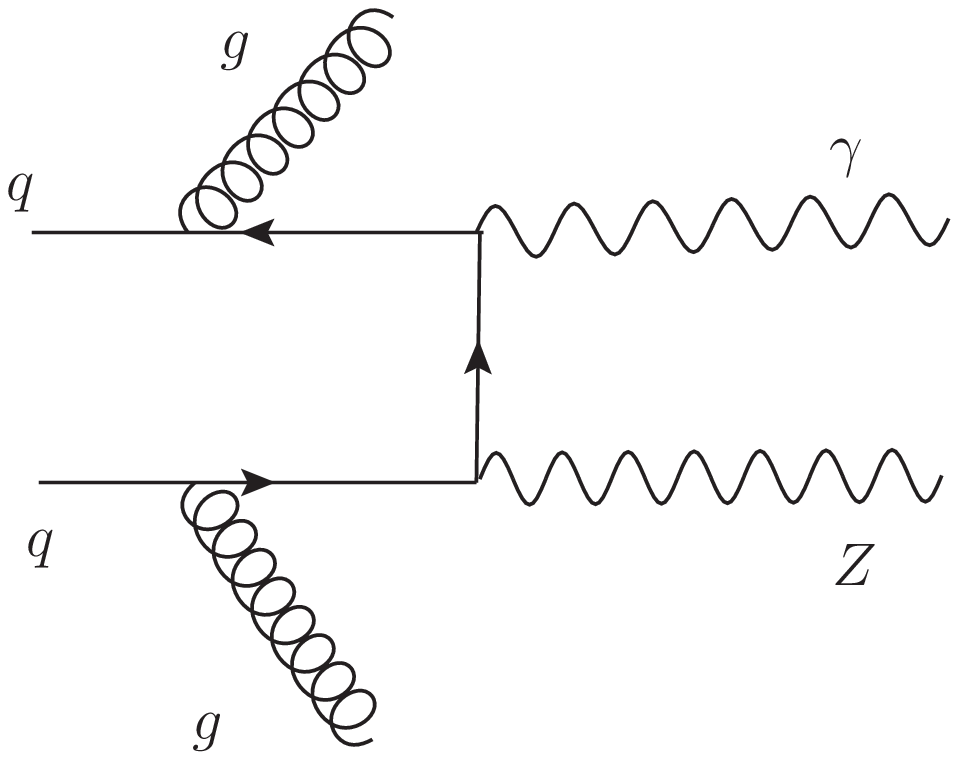}
  \caption{Feynman diagrams of electroweak $Z\gamma jj$ production
    involving VBS subprocesses (bottom left) or non-VBS subprocesses (top left) and of QCD $Z\gamma jj$ production with gluon exchange (top right) or radiation (bottom right).~\cite{jhep:ZyVBS_ATLAS}}
  \label{fig:fey_dia}
  \end{center}
\end{figure}

The Feynman diagrams shown in Figure~\ref{fig:fey_dia} represent the measured signal and the background processes.

\section{Data and Monte Carlo}

ATLAS is one out of the two general purpose experiments at the Large Hadron Collider (LHC) at CERN. It operates with a multipurpose particle detector with a forward–backward
symmetric cylindrical geometry and an almost 4$\pi$ coverage in solid angle, consisting of the inner tracking system, electromagnetic/hadronic calorimeter system and outermost
muon spectrometer system. The dataset used in this analysis is collected by ATLAS in the year of 2012 at the center-of-mass energy of 8 TeV, with an integrated luminosity of 20.2 fb$^{-1}$.
The signal process of electroweak $Z\gamma+jj$ and the irreducible background of QCD induced $Z\gamma+jets$ are both modeled by \textsc{Sherpa}, while the other subleading backgrounds are either
estimated from data (e.g. $Z+jets$) or from Monte Carlo (e.g. $t\bar{t}+\gamma$).

\section{Methodology}

The Fiducial cross section of the $\ell \ell \gamma jj$ channel is measured by
applying a likelihood fit to the spectra of $Z\gamma$-Centrality $\zeta_{Z\gamma}$
simultaneously to a QCD background enriched control region (CR) and a signal enriched signal region (SR),
constraining $Z\gamma jj$-QCD background.
The $Z\gamma$-Centrality is defined as:
\begin{equation}
\label{eqn:centrality}
 \zeta  \equiv \frac{\eta - \bar{\eta}_{jj}}{\Delta \eta_{jj}},
\end{equation}
The CR is used to constain the QCD background and fit to the data, with the theory and experimental systematics fully correlated to SR in the simultaneous fit.
The SR is then used to extract the signal while the background is simultaneously fitted. The other backgrounds which are much less dominant are subtracted in both SR and CR when fitting.

In $\nu \nu \gamma jj$ channel, the fiducial cross section is measured in an aQGC sensitive high $E_T(\gamma)$ region via log-likelihood fit.
The aQGC search region is obtained with a high $E_T(\gamma)$ cut at 150 GeV so as to enhance the sensitivity to the aQGCs and suppress the SM backgrounds.

\section{Result}

As a result, the fiducial cross section are measured in $\ell \ell \gamma jj$ channels with EWK-only in SR and EWK+QCD in both CR and SR. In $\nu \nu \gamma jj$ channel, only EWK-only
crosss section limit is set in a region sensitive to aQGC at 95\% CL as 1.06 fb and 0.99 fb for measured and expected, respectively.

The expected and observed aQGC 1D limits are set on dimension-8 effective field theory (EFT) parameterized operators by performing a one-dimensional profile-likelihood fit with all the
other operator coefficients set to zero, combining both charged lepton and neutrino channels.

Table \ref{tab:cs_summary} shows the measured cross sections in the charged lepton channel and Table~\ref{tab:observedExpected1DLimitsAQGC_comb} shows the 95\% CL intervals on the aQGC parameters.
The measured fiucial cross-sections are consistent with the QCD next-to-leading order prediction by \textsc{VBFNLO}. The aQGC limits are by far the most stringent limits in $Z\gamma+jj$ channel.

\begin{table}[h!]
  \centering
  \begin{tabular}{lccr@{$\,\pm\,$}c@{$\,\pm\,$}lr@{$\,\pm\,$}l}
    \hline
Channel & Phase-space  & Process & \multicolumn{3}{c}{Measured}           & \multicolumn{2}{c}{Predicted} \\
        & region       & type    & \multicolumn{3}{c}{cross-section [fb]} & \multicolumn{2}{c}{cross-section [fb]} \\
    \hline
$Z(\ell^+\ell^-)\gamma jj$ & Search region & EWK      & 1.1 & 0.5\ \text{(stat)}  & 0.4\ \text{(syst)}  & 0.94 &  0.09 \\
$Z(\ell^+\ell^-)\gamma jj$ &Search region & EWK+QCD   & 3.4 & 0.3 \ \text{(stat)} & 0.4 \ \text{(syst)} & 4.0  &  0.4  \\
$Z(\ell^+\ell^-)\gamma jj$ &Control region & EWK+QCD  & 21.9  & 0.9 \ \text{(stat)}   & 1.8 \ \text{(syst)}   & 22.9   &  1.9    \\
    \hline
  \end{tabular}
  \caption{Summary of $Z\gamma jj$ production cross-section measurements in the search and control
     regions for the charged-lepton channel.~\cite{jhep:ZyVBS_ATLAS}}
  \label{tab:cs_summary}
\end{table}
\clearpage

\begin{table}
\small
\begin{center}
\begin{tabular}{ccccc}
        \hline
              & 95\% CL intervals  & Measured [TeV$^{-4}$] & Expected [TeV$^{-4}$] & $\Lambda_{\mathrm{FF}}$ [TeV] \\ \hline
\multirow{7}{*}{$n=0$} & $f_{T9}/\Lambda^4$ & $[-4.1, 4.2]\times10^3$  & $[-2.9, 3.0]\times10^3$ & \\
              & $f_{T8}/\Lambda^4$ & $[-1.9, 2.1]\times10^3$  & $[-1.2, 1.7]\times10^3$ &  \\
              & $f_{T0}/\Lambda^4$ & $[-1.9, 1.6]\times10^1$  & $[-1.6, 1.3]\times10^1$ & \\
              & $f_{M0}/\Lambda^4$ & $[-1.6, 1.8]\times10^2$  & $[-1.4, 1.5]\times10^2$ & \\
              & $f_{M1}/\Lambda^4$ & $[-3.5, 3.4]\times10^2$  & $[-3.0, 2.9]\times10^2$ & \\
              & $f_{M2}/\Lambda^4$ & $[-8.9, 8.9]\times10^2$  & $[-7.5, 7.5]\times10^2$ & \\
              & $f_{M3}/\Lambda^4$ & $[-1.7, 1.7]\times10^3$  & $[-1.4, 1.4]\times10^3$ & \\ \hline
\multirow{5}{*}{$n=2$} & $f_{T9}/\Lambda^4$ & $[-6.9, 6.9]\times10^4$  & $[-5.4, 5.3]\times10^4$ & 0.7\\
              & $f_{T8}/\Lambda^4$ & $[-3.4, 3.3]\times10^4$  & $[-2.6, 2.5]\times10^4$ & 0.7 \\
              & $f_{T0}/\Lambda^4$ & $[-7.2, 6.1]\times10^1$  & $[-6.1, 5.0]\times10^1$ & 1.7 \\
              & $f_{M0}/\Lambda^4$ & $[-1.0, 1.0]\times10^3$  & $[-8.8, 8.8]\times10^2$ & 1.0\\
              & $f_{M1}/\Lambda^4$ & $[-1.6, 1.7]\times10^3$  & $[-1.4, 1.4]\times10^3$ & 1.2 \\
              & $f_{M2}/\Lambda^4$ & $[-1.1, 1.1]\times10^4$  & $[-9.2, 9.6]\times10^3$ & 0.7 \\
              & $f_{M3}/\Lambda^4$ & $[-1.6, 1.6]\times10^4$  & $[-1.4, 1.3]\times10^4$ & 0.8 \\ \hline
\end{tabular}
\caption{Measured and expected one-dimensional 95\% confidence level intervals
  on aQGC parameters (in the \textsc{vbfnlo} formalism)
  using the combination of all $Z\gamma jj$ channels (charged-lepton and neutrino).
  The Form-Factor (FF) exponent $n=0$ entries correspond to an infinite FF scale and therefore result in
  non-unitarized 95\% CL inntervals. FF exponent $n=2$ confidence intervals preserve unitarity with individual form-factor
  scales as shown in the last column for each dimension-8 operator.~\cite{jhep:ZyVBS_ATLAS, bib:ZyNLO}}
\end{center}
\label{tab:observedExpected1DLimitsAQGC_comb}
\end{table}

\end{document}